\documentclass[letter,12pt]{article}
\usepackage{graphicx,amssymb,amsmath,epstopdf,color,hyperref}

\def\beq{\begin{equation}}
\def\eeq{\end{equation}}
\def\bea{\begin{eqnarray}}
\def\eea{\end{eqnarray}}

\def\gev{\, {\rm GeV}}

\newcommand{\gsim}{\lower.7ex\hbox{$\;\stackrel{\textstyle>}{\sim}\;$}}
\newcommand{\lsim}{\lower.7ex\hbox{$\;\stackrel{\textstyle<}{\sim}\;$}}

\def\xsection#1{\section{#1}}

\begin{document}

\begin{flushright}
November 21, 2019
\end{flushright}

\vspace{.50in}

\noindent
\begin{center}

{\bf\large The Once and Present Standard Model \\ of Elementary Particle Physics}

\vspace{1.0cm}
{ James D. Wells}

{\it Leinweber Center for Theoretical Physics \\
University of Michigan, Ann Arbor, MI 48109 USA} \\

\end{center}

\vspace{1cm}

\noindent
{\it Abstract:} 
There are many theories that have resided these last fifty years within the  hazy mist we have been calling the Standard Model (SM) of elementary particles. An attempt is made here to construct a coherent description of the  SM today, because only precisely articulated theories can be targeted for annihilation, corroboration, and alteration. To this end it is useful to categorize the facts, mysteries and myths that together build a single conception of the SM. For example, it is argued that constructing a myth for how neutrinos obtain mass is useful for progress. We also advocate for interpreting the cosmological constant, dark matter, baryogenesis, and inflation as four ``mysteries of the cosmos" that are indeterminate regarding new particles or interactions, despite a multitude of available particle explanations.   Some history of the ever-changing SM is also presented to remind us that today's SM is not our parents' SM, nor will it likely be our children's SM.

\vspace{2cm}

\begin{center}
{ To appear in J.D. Wells, {\it Discovery Beyond \\ the Standard Model of Elementary Particle Physics.} Springer, 2019}
\end{center}

\vfill\eject

\tableofcontents

\xsection{Introduction}

We have known that neutrinos have mass for over two decades, and we had theoretically and experimentally built support for the case that neutrinos had mass for several decades prior to that. Yet, we continue to say phrases like ``neutrinos are massless in the Standard Model"\footnote{For example, ``In the Standard Model (SM) of elementary particle physics, neutrinos are massless particles"~\cite{Winter:2010hb}. And, ``Some considered [the Higgs boson] as the last brick in the construction of the Standard Model. It is not, since in the Standard Model neutrinos have no mass..."~\cite{Valle:2017fwa}. And, ``The standard model of particle physics says neutrinos should be massless, but experiments have shown that they have a small but nonzero mass --- the subject of the 2015 Nobel Prize in Physics"~\cite{Hill:2019}. There are many more such quotes throughout the literature and presented in talks.}. This is certainly not out of ignorance, since it is being said by outstanding scientists who are not confused.
 An underlying reason for this is because we as a community  have never really confronted precisely what we mean when we say ``Standard Model." Does it mean what physicists envisioned it to mean in 1974, and thus is a static definition tied to predilections at the beginning of modern particle physics (no neutrino masses, no third generation, shakiness on Higgs boson, rising premature acceptance of grand unification, etc.)? Or, is ``Standard Model" a dynamic name that is equivalent to ``current standard theory" of particle physics, which continually updates itself over time to incorporate the community's current view of the most favored and agreed-upon description of elementary particle physics (including neutrino masses, etc.)? If it is  the old static former definition, then it is uninteresting to use the phrase ``Standard Model" ever again, except in nostalgic history books, and if it is the new dynamic latter definition then we should not speak of neutrino masses being beyond the Standard Model since it implies we are unable as a community to incorporate that fact into a theoretical structure. Of course, the lack of an agreed upon module for incorporating neutrino masses is at the origin of this confusion with the word, and should be acknowledged. Nevertheless, it is time we cease using the phrase ``Standard Model" for a theory we know to  be incorrect. 

In this article, we advocate for the more useful definition of ``Standard Model" (SM), meaning the ``current standard theory" of elementary particle physics. In this sense there have been ``Standard Models"  well before there were quarks and leptons.  In the early 20th century it was thought to be entirely made of electrons and protons, and then neutrons were added, and then an explosion of other discoveries happened (leptons, quarks, etc.) bringing us to the modern age.

One can reasonably argue that the modern age of particle physics started in 1974 with the discovery of charm. This ``November revolution"~\cite{Close:2004mj,Bjorken:1984} was the final offensive that forced all competent resistance to surrender to the $SU(3)_c\times SU(2)_L\times U(1)_Y$ renormalizable gauge theory with elementary quarks and leptons. Only the Higgs boson thereafter faced significant opposition by competent experts before winning the day in 2012~\cite{Wells:2018nwj}. The SM in place in 1974, however, is not the SM that is in place now. By this I do not mean only that the gauge couplings have been measured better and the Higgs boson has been experimentally confirmed. In other words, points in continuous parameter space measured continually better but not violating any {\it a priori} assumed theory structure do not constitute a dismissal of one SM for another. Nor is the confirmation of an elementary particle that had not yet been seen but was {\it part of the defined theory} a strong reason to say that the original SM is not the same as today's SM. What is valid to say is when there is a qualitatively new phenomenon that was not anticipated in the original SM formulation, and has been corroborated by experiment and understood theoretically, then the old SM must be replaced by a new SM, and it should be recognized as such.

One is therefore drawn to attaching dates to the SM, just as one attaches dates to substantial revisions of a computer programming language (Fortran 77, Fortran 90, etc.). The SM in 1974, which one could call SM-1974, is certainly not the SM of today, SM-2019. They differ by incorporation of third generation, and the acceptance of neutrino masses, and perhaps other ways in which SM-2019 may be defined more precisely. Further defining features may include assumptions on non-renormalizable operators, grand unification, energy domain of applicability, dark matter, $\theta_{\rm QCD}$ value for strong CP, etc. We discuss these issues more below, with a major goal of stating what is required to define a SM and what is our SM today.

There is value in articulating very precisely what is the standard theory of some particular domain, such as the SM of particle physics. Precise articulation increases understanding and precision of claims made; it enables clarity on progress in theory development; it forces one to confront areas of ambiguity when equally minimal/simple sectors of the theory compete to be incorporated within the SM (particularly relevant for neutrino physics and dark matter); and, it provides a very clear target for organizing attempts to kill the theory. Fuzzy theories are harder to falsify, and when a theory is susceptible to eliminating its fuzziness, it should do so. The SM is most certainly capable of tightening up its definition, especially within the sector of neutrino physics, as a useful step toward killing it. That is one of our primary aims below.

\section{Facts, Mysteries and Myths}

When working out a theoretical picture for natural phenomena it is useful to separate out facts, mysteries and myths. I will refer to facts as data and conceptual categories that are not presently under question by anyone in the community. For example, the existences of electrons, muons, and $W^\pm$ bosons are facts. The realization that neutrinos have mass is another fact. I am ignoring some philosophical subtleties with this free use of the word ``facts" but it is adequate for our working purposes here. 

Mysteries are questions that we think are important which we cannot answer at the present but we believe there is more information ``out there" that can elucidate them in part or in whole. This would include the plausible discovery of new particles or new interactions that directly answer the questions, or at least elucidate them at a deeper level, or it even could include understanding new concepts that remove the question from further interest. Examples of mysteries today are, why is the weak scale so far below the Planck scale? Does the converging behavior of gauge couplings at the high scale signify unification of the forces? How do quantum mechanics and general relativity co-exist peacefully in a unified theoretical structure? What is the origin of three generations of quarks and leptons? What is the origin of four-dimensional spacetime? What is the origin of the phenomena that supports the existence of dark matter? Why is there more matter than anti-matter? There are many additional such questions. The SM has no strong claim to the answers to those questions, yet ideas exist that appear to be satisfactory theoretically and empirically. In sec.~\ref{sec:cosmosmysteries} I will focus on the cosmological questions as the key mysteries of the SM today.

Myths are  mysteries that have been answered concretely, yet the answers have a reasonable probability of being wrong, incomplete or naive. Nevertheless, the community accepts the myths for several useful reasons: they are easy to understand, they elevate the mystery in the consciousness of the community, and perhaps most importantly they enable interpretation of phenomena in a well articulated manner. Before it was discovered in 2012, the Higgs boson fully qualified as a myth within the SM. Some might even say that it continues to be myth if one speaks of the Higgs boson as purely the SM-conceived Higgs boson without any possible deviations, but that would be stretching the myth concept. Here, we can put Higgs boson in the fact column now, making it part of the firm SM, and move on. However, myths with respect to the SM still exist. For example, below I will advocate a specific articulated scenario as a useful myth of how neutrinos obtain their mass.

Therefore, we can define the SM as possessing facts (its known particles and interactions), mysteries (e.g., the origin of matter asymmetry), and myths (e.g., how neutrinos get their masses). Beyond the SM (BSM) physics primarily concerns itself with articulating the mysteries and myths, developing concrete answers to the mysteries, and identifying phenomena and experiment that can shed light on them.

 Articulating a tentative definition of a more complete minimal SM was also contained in other works over the last few years~\cite{Davoudiasl:2004be,Ballesteros:2016xej}. In our language, those works proposed a much richer myth structure than what will be advocated here. These were very useful and laudable exercises. The SM-2004 theory presented in~\cite{Davoudiasl:2004be}, for example, was economical and certainly valid and intriguing at the time. However, it may not be the most minimal in current eyes, and recent experimental developments have put stress on some of the ideas and perhaps point to a different standard choice to make for the SM's relationship to cosmology in particular. That is partly why I prefer putting the many cosmological conundrums in the mysteries column of the SM rather than proposing specific myths for their resolutions. Such considerations will be explained in more detail below. 

\xsection{Requirements for a Standard Theory} 

A well known challenge of science is called the ``underdetermination problem", which suggests that there is usually more than one theory that can accommodate a set of experimental results~\cite{Laudan:1991}. Our experience within particle physics suggests to us an even stronger claim, which we can call the ``infinite underdetermination problem" (IUP), which states that there are an infinite number of theories that can accommodate a finite number of imprecise observables. By ``imprecise observable" we mean measurements that lead to a determination of an observable with non-zero error (for cross-sections, branching ratios, etc.). 

The evidence for IUP is compelling. There are an infinite number of sets of higher-order operators (i.e., arising from an infinite number of unique high energy theories) that decouple from low energy phenomena, yet give tiny shifts in values below error bars of current measurements. Among this infinite number of theories, there is usually a theory class that rises to the top among community researchers because it features a number of desired virtues, such as simplicity, calculability, consilience, unification and of course consistency with experiment. The theory that rises to the top is the ``standard theory" for the domain in question.

For particle physics, the standard theory has been called the Standard Model (SM). As we discussed above, we will attach  the name SM to the current standard theory of elementary particles and their interactions, as opposed to viewing it as a name for the standard theory that was in place in the early 1970s. Yet, we must ask, does the SM satisfy all the requirements of a standard theory, and if not, what more must we specify?

The standard theory must be 1) a precisely articulated physics theory that is 2) recognized to be the leading theory among the community of scholars. Most would agree that the SM satisfies 2), which is ironic because one usually wants to know what they are voting for, which is impossible to do because the SM comes up short on 1).

To specify what the SM is, one must first decide what we are asking of it as a theory. A physics theory ($T$) is a set of rules  that maps input parameters $\{ \xi_i\}$ to experimental observables $\{ {\cal O}_k\}$ over an agreed-upon target domain $\Delta_{\cal O}$ of possible observables. Symbolically we can say
\beq
T: \{ \xi_i\}\longrightarrow \{ {\cal O}_k\}~{\rm where~all}~ {\cal O}_k\in \Delta_{\cal O}.\nonumber
\eeq
The definition of an observable can be subtle, since we are used to observables being defined within the theory framework. For example, the cross-section $e^+e^-\to t\bar t$ requires us to have a conception of what an electron is and what a top quark is, which is only provided by the theory itself. Thus, there is some inevitable circularity in the definition of a theory, but that circularity is put under the stringent test of experimental and observational self-consistency. There are ways to reduce that circularity, but such efforts would not be of much practical value in our discussion here. We assume here that there is an intuitive, yet ultimately precise, understanding of what observables are, and what a domain for observables can be.

A standard theory of elementary particles should by definition be a theory of all the putative elementary particles (indivisible) and their interactions. On the surface that is the easy part. One could say that the full set of elementary particles are the fermions (quarks, leptons, and neutrinos), the force carrier vector bosons (photon, $W^\pm$, $Z^0$, and graviton) and the Higgs boson. Now, regarding the particle content of the SM, if we are content with such imprecision (we should not be), we end up lazily not recognizing many things that have happened over the years since the original SM's birth in 1974. More will be said below about tightening the theory discussion of particle content.

Regarding the target domain of observables ($\Delta_{\cal O}$), that is also subtle, which is related to the subtlety of defining observables discussed above. If we believe that we have a theory of all elementary particles and all the forces that apply, then in principle we have a ``final theory" since everything takes place ultimately at the elementary particle level. Thus,  we should in principle be able to not only predict the lifetime of the top quark, but we should also be able to predict the next earthquake.  Yet, earthquakes are not within the observables target domain $\Delta_{\cal O}$ for a standard theory of particle physics, for reasons that are well-known and do not need to be reviewed here. Such examples are not the origin of the target domain subtlety, but rather what energy ranges do we assume the theory to be valid, and in general, what conditions must hold within the target domain of observables.

To this end, we can define the SM more precisely to be a theory ($SU(3)\times SU(2)\times U(1)_Y$ gauge theory) of elementary particle content (quarks, leptons, neutrinos and force carriers) with a parameter space of inputs (gauge couplings, Yukawa couplings, etc.)\ that enable unambiguous computations of  decay lifetime observables ($\Gamma_i \subset \Delta_{\cal O}$) and interaction cross-section observables ($\sigma_{ab\ldots}\subset \Delta_{\cal O}$) in non-extreme gravitational environments (further restriction on $\Delta_{\cal O}$), where computed results are all consistent within experimental uncertainty for at least one point in the parameter space. By non-extreme gravitational environments we mean when momentum transfers in parton-level collisions are small compared to the Planck mass of $\sim 10^{18}\gev$, or stated more generally, when  the uncertainty of our understanding of strong gravity does not obviously get in the way of computability\footnote{It is generally held that any particle physics theory based on standard quantum field theory will break down in extreme gravity environments, which is one of the motivations for pursuing deeper string theory descriptions for that domain. It is also why we restrict our discussion to energies well below the Planck scale.}.

With this more comprehensive definition of what we require out of a standard theory of particle physics, we investigate how the SM has changed since its inception and point out how our current usage of the word ``Standard Model" is foggy, and we present a suggestion for making it more precise in a way that satisfies the demands of a standard theory. We require that our more precisely defined SM be within the foggy domain of what is currently meant by the SM, and that it  have a strong prospect for being falsified by near-future investigations. And lastly, it should be noted, as with all standard theories in any domain, acceptance of it as the standard theory in no way commits a physicist to believe that it is the ``right" or ``correct" theory that will remain valid forever in the face of all future theoretical and experimental stresses put to it over time. It merely is designated as the standard theory among the infinite number of currently viable theories that has maximal theory virtues valued by the community at this moment in time. In other words, defining the SM more precisely in no way should be interpreted as heightened arrogance that we know exactly what nature has chosen. Rather, it is a tool through which we track our understanding through time and create firm targets to attack theoretically and experimentally.

\xsection{Historical Progression of the Standard Model}

In the previous section it was stated that a precise statement on the particle content is required for the SM. It should be kept in mind that there is a difference between when a particle entered the standard expectations of the community, and thus was incorporated within the SM, and when the particle was actually confirmed by experiment. In other words, when tracking the evolution of the SM (i.e., the accepted standard theory of expectations) with respect to particle content, one should focus  more on when particles were expected and not on when they were discovered.

For example, the Higgs boson is viewed as one of the most revolutionary discoveries in particle physics in the last fifty years, and rightly so. However, it has been a part of the SM since the beginning. What made it so momentous is that it is a qualitatively new type of elementary particle -- a spin-zero boson -- that had never been discovered before. As such, it was highly controversial and many competent experts had strong suspicions that it {\it could not exist} even up to the time of its discovery~\cite{Wells:2018nwj}. Nevertheless, it was already part of the SM as an accepted myth --- the expectations within the standard theory. Thus, it is not the Higgs boson that  has been the source of change over the years to upend one SM in favor of a new SM, despite the extraordinary impact its confirmation discovery has made on science.

\noindent
{\it Fermion generations}

What has changed is our conception of fermion generations and our conception of the neutrino sector. Let us look at the fermion generation question first. When the SM emerged out of the Glashow-Salam-Weinberg model of electroweak interactions, complete with spontaneous symmetry breaking from the Higgs boson, and combined with the new understanding of quarks and QCD, there was not initially an understanding that there were more than two generations of fermions. 

The understanding of the need for three generations came from theory and experiment. In theory work, it was suggested correctly by Kobayashi and Maskawa~\cite{Kobayashi:1973fv} that a third generation of fermions is needed in order to accommodate CP violation in the kaon system if its origin is through weak interactions. Two generations would not enable complex phase (i.e., CP violation) in what we today call the Cabibbo-Kobayashi-Maskawa (CKM) matrix, but three or more generations do. Later, the third generation was established by experiment over a range of lepton and hadron collider experiments that discovered the tau lepton, bottom quark, tao neutrino, and top quark over a twenty year period, ending in the top's discovery in 1995. 

\noindent
{\it The original Standard Model (SM-1974)}

In the original SM, which we can perhaps call SM-1974, there were only two generations of fermions, with only the charm quark missing. In J.D.~Bjorken's 1984 recollections of those early times he described a great uncertainty and fogginess about what really was the underlying physics. He recalled that John Ellis had given a summary talk at an international particle physics conference in London in mid-1974 summarizing all the theory interpretations of the data. As Bjorken remembered it,  
\begin{quote}
Ellis' catalog well reflected the state of theoretical confusion and general disarray in trying to interpret $e^+e^-$ data. But in the midst of all of this was a talk by John Iliopoulos .... With passionate zealotry, he laid out with great accuracy what we call the standard model. Everything was there: proton decay, charm, the GIM mechanism of course, QCD, the $SU(2)\times U(1)$ electroweak theory, $SU(5)$ grand unification, Higgs, etc. It was all presented with absolute conviction and sounded at the time just a little mad, at least to me (I am a conservative). So at London the pressure to search for charm was there. But even so this was immersed in a rather large degree of confusion.\cite{Bjorken:1984}
\end{quote}
Bjorken proceeds to describe the confusion regarding experimental attempts to confirm the existence of the second generation charm quark, but then the revolution happened. In Bjorken's words:
\begin{quote}
That brings us up then to November 1974. The stage was really set. The balance had changed, and the November revolution just set everything into motion toward the standard model that we have now. Most high energy physicists will probably remember where they were when they first heard about the psi [$J/\psi$ charm meson]. It is like the moon landing, Pearl Harbor or the Kennedy assasination. I was home and it was dinner hour. Burt Richter called me up and told me the basic parameters over the phone. He said three GeV. I said three GeV per beam, right? He said no, three GeV in the center of mass. I couldn't believe such a crazy thing was so low in mass, was so narrow, and had such a high peak cross-section. It was sensational.~\cite{Bjorken:1984}
\end{quote}

Indeed it was sensational for all the physicists as described by Bjorken here and by others elsewhere (see also~\cite{Close:2004mj}).

\noindent
{\it Unification and the Standard Model}

It is interesting to be reminded by the first Bjorken quote above that in the particle physics community's eyes, from the mid 1970s to early 1980s,  $SU(5)$ unification, with its generic prediction of proton decay, was extraordinarily compelling. One might even be tempted to put $SU(5)$ unification within the SM-1974 definition, which was then dropped later from the ambient SM mindset after initial experiments looking for proton decay in the early 1980s did not find it. That is a refinement that would be interesting to describe further, but for our purposes here we wish to merely describe the birth of the standard theory, give a feel for how often and substantively the views of the standard theory have changed over the years, and define a current SM that reflects community sentiment. 

\noindent
{\it Neutrino masses: disbelief}

Coming back to the original SM theory of 1974, we note that it only had two generations of fermions, except for the latent hint from Kobayashi-Maskawa's 1973 work~\cite{Kobayashi:1973fv} that CP violation can be achieved in weak interactions by a third generation of quarks. In addition to uncertainty about the number of generations assumed in the SM, there was little questioning that the neutrino masses were most likely zero, and thus zero mass neutrinos were a cornerstone of the SM definition. The SM did not have, therefore, right-handed neutrinos $\nu_R$ in the spectrum, nor did it recognize or allow for the possibility of the dimension-five Weinberg operator of left-handed neutrinos and Higgs boson $(LH)^2/\Lambda$, where $\Lambda$ is required to be substantially higher than the weak scale due to the extreme lightness of the neutrino masses compared to other known masses. It is not as though they could not conceive of neutrinos having the possibility of being massive. They did (see, e.g.,~\cite{Bilenky:1977du,Marciano:1977wx,Lee:1977qz}). It was merely the case of having no compelling evidence for neutrino masses, yet having evidence that if they did exist they had to be many orders of magnitude below the mass of all other known elementary particles. This suggested that it was better to be zero than bizarrely and unexplainably low.

It was not until 1979/1980 that the possibility of neutrinos having mass started gaining widespread community traction. This was the time when the neutrino seesaw became widely known and appreciated within the community, which gave a good reason why neutrino masses could be naturally very tiny compared to the other leptons. Experimental searches were also underway, and first signs of neutrino oscillations, which implied neutrino masses, became evident in Ray Davis's pioneering Homestake experiment in the late 1960s and early 1970s which lead to the ``solar neutrino problem"~\cite{Davis:1994jw}. 

Nevertheless, the ``solar neutrino problem" was viewed as inconclusive since it did not detect all the other neutrinos into which it could have oscillated, and there was question as to how well we understood the sun's complex internal processes, etc.~\cite{Pinch:1986}. As Ray Davis reported, ``My opinion in the early years was that something was wrong with the standard solar model; many physicists thought there was something wrong with my experiment"~\cite{Davis:2002}. For example, Trimble and Reines's 1973 review on the solar neutrino problem states: ``The critical problem is to determine whether the discrepancy is due to faulty astronomy, faulty physics, or faulty chemistry"~\cite{Trimble:1973ca}.
Nevertheless, the theory and experimental progress that proceeded led the community  from the early 1970s to the early 1990s to adiabatically come around to the expectation, not just the possibility, that neutrinos had mass. 

To demonstrate the widely held belief even in the 1980s that neutrinos were massless we can refer to Cheng and Li's {\it Gauge theory of elementary particle physics} published in 1984~\cite{ChengLi:1984}, which was one of the most widely read advanced particle physics textbooks. It had this to say about neutrino masses:
\begin{quote}
We have seen that the standard theory [now with 3 generations] gives a natural explanation for the presence of the Cabibbo angle and CP phases in quark charged currents. Similarly the same theory helps us to understand the absence of such features in the lepton sector; the masslessness of neutrinos implies that these mixings are physically unobservable.~\cite{ChengLi:1984}
\end{quote}
and
\begin{quote}
We have already mentioned in \S 11.3 that the reason why there are no Cabibbo-like mixing angles in the lepton sector of the standard electroweak theory is neutrino mass degeneracy (i.e.\ all $\nu$s have the same mass --- zero). This degeneracy means that there is no need to diagonalize the neutrino mass matrix (in fact no mass mass matrix to begin with).~\cite{ChengLi:1984}
\end{quote}
It is fair to say that there was widespread skepticism and even disbelief of neutrino masses even into the 1980s.

\noindent
{\it Neutrino masses: rising belief}
 
Throughout the late 1980s and early 1990s the mainstream perspective on neutrino masses shifted. For example, 
the Particle Data Group bi-yearly updates transitioned from  statements like, ``If one considers the possibility of nonzero masses for neutrinos\ldots" in 1992~\cite{PDG:1992} to something much more definitive in 1994 about the community's expectation that neutrinos have mass:
\begin{quote}
The theoretical perspective concerning neutrino masses has changed considerably over the past 20 years. Before that time, a standard view was that there was no theoretical reason for neutrinos to have masses.... Indeed, even in the literature of the 1970's, one will often find statements asserting that in the standard $SU(2)\times U(1)$ electroweak theory ... the known ... neutrinos are massless. In contrast, in the modern theoretical view ... small but nonzero neutrino masses are expected on general grounds.~\cite{PDG:1994}
\end{quote}

Certainly by the early 1990s the community was firmly behind the proposition that neutrinos had mass and it was part of the standard theory.  The pressure that took the community from no masses to belief in neutrino masses was due to many factors, including ``neutrino anomalies" among several experiments and the rising new theory perspectives of string theories, grand unified theories, and supersymmetry. These theory perspectives liked neutrino masses and were gaining indirect experimental support from LEP precision measurements of the gauge couplings that pointed toward supersymmetric unification, further supporting string theory ideas and by consequence the other intuitions that came with it, including neutrino masses (especially through $E_6$ unification and its subgroup $SO(10)$). Despite the standard theory never being supersymmetric, nor being described directly as a string theory,  their influence extended to the low-energy standard model theory expectations, and thus the expectations of neutrino masses were solidified both experimentally and theoretically, and anomalies began to be interpreted as evidence for mass. For example, the arguments delivered by Robert Shrock in the PDG in 1994~\cite{PDG:1994} for why non-zero neutrino masses are expected are primarily from a string theory perspective, which reflects the sentiments of the writer who, in his community responsibility as PDG contributor, is presumably summarizing widely held viewpoints. In any event, it is rather safe to say that SM-1990s was a theory with neutrino masses.

Neutrino masses were experimentally beyond reproach finally by 1998. That is when Super-Kamiokande firmly established a self-consistent and comprehensive picture of neutrino oscillations~\cite{Fukuda:1998mi}.  Ever since there has been no question about its required presence within the SM. The mass differences and mixing angles have been measured with increasing precision over the years since. A new push to measure the neutrino sector even more precisely is underway, including the many current and future flagship programs at Fermilab~\cite{Valle:2017fwa}.

\noindent
{\it Neutrino masses: theories}

Despite all of this attention on neutrino physics, there is no clear agreement of what the neutrino sector is within the SM. There are many possibilities. One possibility is to add a right-handed neutrino and then add the Yukawa operator 
\beq
\label{eq:L1}
{\cal L}_1\subset y_\nu \bar L H\nu_R~~~~{\rm (Dirac~neutrinos)}.\nonumber
\eeq
These Dirac neutrinos are then given mass according to $m_\nu=y_\nu \langle H\rangle$. Or, it could be defined without adding any new particles, and the masses are generated by the dimension-five Weinberg operator 
\beq
{\cal L}_2\subset \frac{1}{\Lambda}(\bar L H)(\bar L H)~~~~{\rm (Majorana~neutrinos)}\nonumber
\eeq
where $\Lambda$ is a scale much higher than the electroweak scale in order to give tiny Majorana neutrino masses, which are given by $m_\nu=\langle H\rangle^2/\Lambda$. Or, yet further, one could supplement ${\cal L}_1$ with a right-handed neutrino Majorana mass to give
\beq
{\cal L}_3\subset y_\nu \bar L H\nu_R+M\nu_R\nu_R~~~~{\rm (seesaw)}\nonumber
\eeq
which, if $M\gg y_\nu\langle H\rangle$, yields light neutrino masses of mostly left-handed composition with mass $m_\nu\sim (y_\nu \langle H\rangle)^2/\Lambda$. This is the famous seesaw mechanism for generating tiny neutrino masses even if $y_\nu\langle H\rangle$ is on order of other leptons and quarks in the theory~\cite{Minkowski:1977sc,GellMann:1980vs,Yanagida:1979as,Mohapatra:1979ia,Schechter:1980gr}.

Thus, we have three theories to consider for neutrino masses, all of which co-exist in a fuzzy superposition of what the community refers to when they say SM. In order to define SM-2019 precisely we need to choose which of the many theories is to be designated the standard theory. The two simplest are ${\cal L}_1$ and ${\cal L}_2$, so we eliminate ${\cal L}_3$ from the running. ${\cal L}_1$ has the advantage of introducing only three $\nu_R$'s along with a  corresponding Yukawa coupling matrix $y_\nu$ whose entries are additional input parameters to the SM. ${\cal L}_2$ has less fields (i.e., no $\nu_R$'s) and the same number of input parameters associated with the coefficient matrix to the Weinberg operator. However, it is not a viable theory across the full energy range of interaction strengths that are not gravitationally strongly coupled. For example, scatterings of neutrinos with energies well above $\Lambda$ yet well below $M_{\rm Pl}$ ($\Lambda\ll E_\nu\ll M_{\rm Pl}$) are generically expected to be altered, whereas the interactions added in ${\cal L}_1$ have no immediate worries for where they will break down and so are well-behaved and calculable all the way up to near $M_{\rm Pl}$. 

\noindent
{\it Neutrino masses within the SM}

For the reasons stated above, an excellent candidate for SM-2019 --- the standard theory of elementary particle physics up to $M_{\rm Pl}$ that includes neutrino masses --- is one that incorporates neutrino masses by introducing three $\nu_R$'s and Yukawa couplings of it to the left-handed doublets $L$ and Higgs boson, as described by ${\cal L}_1$ in eq.~\ref{eq:L1}.

We can go one step further by considering an important question within neutrino physics regarding whether the neutrino mass eigenstates are organized in a normal hierarchy (NH) or inverted hierarchy (IH). The normal hierarchy suggests that the neutrinos that are most matched with the leptons (i.e., flavor eigenstate overlaps) have the same mass hierarchy as the leptons. In other words, is the heaviest neutrino most overlapping with $\nu_\tau$, and is the next massive neutrino most overlapping with $\nu_\mu$, and is the lightest neutrino most overlapping with $\nu_e$? Such an expected ``normal hierarchy" is consistent with all neutrino data. But another possibility is consistent with the data, which has the mass hierarchy inverted in a topologically disconnected region of the parameter space. Knowing which island of parameter space is the correct island, NH or IH, has profound implications for flavor model building~\cite{Altarelli:2010gt} and for prospects to find BSM signals in the neutrino sector (see, e.g.,~\cite{Fogli:2004ff,DellOro:2016tmg,Lei:2019nma}). Thus, it is justifiable to specify one of the islands for SM-2019, while the other is relegated to a BSM possibility.  We choose NH. The justification for NH in SM-2019 springs from two additional reasons beyond what we have already discussed. First, it follows the standard hierarchy that we have learned from quarks. Second, although somewhat controversial, the data may already be giving a slight preference for NH according to recent analyses~\cite{Simpson:2017qvj,Schwetz:2017fey,Caldwell:2017mqu,Long:2017dru,Gariazzo:2018pei,Heavens:2018adv}, and thus may have more empirical claim to be  the choice of SM-2019. 

Of course, a precise description of the neutrino sector cannot be compelling at this point, and therefore it must be introduced not as a fact but as a useful myth, whose implications can be compared with experiment and discovery progress can be tracked, as we will discuss next.

\noindent
{\it Challenging the SM theory of neutrino masses}

One of the advantages of a unambiguous designation for neutrino physics, such as that provided by eq.~\ref{eq:L1} and NH, is that theorists and experimentalists can ask how to falsify and test the theory. And as discussed previous~\cite{Wells:2019zrj} the best way to make progress is to make motivated physics theories for physics beyond the Standard Model  that give experimental predictions that are not within the realm of possibility for the SM. BSM theories with respect to SM-2019 predict new possible phenomena for supernova neutrinos, neutrino oscillation experiments, and neutrino-less double beta decay ($0\nu\beta\beta$) experiments. The hard pursuit of these phenomena is the best way to crack SM-2019 on the path to a qualitatively new SM. SM-2019 retains lepton number conservation at the perturbative level and thus new signals of FCNC in the lepton sector, such as $\mu\to e\gamma$ would signify a breakdown. Discoveries of any new particles or interactions in general would falsify SM-2019. Thus, after fulfilling the pre-requisite of actually defining what the SM means, there is a large class of experiments that could falsify SM-2019 unambiguously in the near future within and outside the world of neutrino experiments. 

In addition to experimental pressure that can be placed on SM-2019 there is much theory pressure to apply. For example, how viable or ``likely" is it that dimensionless Yukawa couplings associated with neutrinos can be so tiny, $y_\nu < 10^{-12}$? Do UV complete theories, such as string theories, allow for such tiny couplings without other accompanying low-scale phenomena predicted? Within SM-2019 there is a conserved lepton number global symmetry which could be broken by adding Majorana mass terms at the renormalizable level for $\nu_R$. Since SM-2019 does not allow these new terms, how stable is that assumption to our attempts to incorporate particle physics with gravity, where it has been conjectured that global symmetries cannot be invoked to prevent otherwise expected terms~\cite{Banks:2010zn,Harlow:2018tng}? What BSM ideas would make the neutrino mass hierarchies and mixing angles more theoretically appealing, and what experimental or observational consequences do these interesting BSM ideas have? All of these questions are on the table, and all may have resolutions to be found in the coming years.

\section{Four Mysteries of the Cosmos}
\label{sec:cosmosmysteries}

Other areas that could conceivably be more precisely well defined in addition to neutrino physics in order to complete our careful designation of the standard theory of particle physics could include new states and interactions that explain the cosmological constant, dark matter, the mechanism that accounts for the preponderance of matter over antimatter, and the mechanism that carried out inflation.

There is no good explanation yet of the cosmological constant. On the other hand, the other three  ``mysteries of the cosmos" (dark matter, inflation, baryogenesis)  have something in common: there are a vast number of adequate, mutually exclusive and even qualitatively different ideas to explain each. Unlike neutrino physics, which has less than a handful of well-disciplined simple theoretical structures to explain their masses and mixings, those three mysteries of the cosmos  have almost no practical bound in the number of ``good ideas" to account for them. A variant on a common aphorism applicable to circumstances like this might say that when you have dozens of mutually exclusive ideas for why something should be true, it means you have no good idea. One should face this fact. There simply cannot be any credible standard theory choice for any of these four mysteries of the cosmos. There is no Secretariat in these races. Every horse is a million to one.

What to do with these {\it mysteria scientiae}? We do as we do with any deep mystery and humbly say ``we do not know." All ideas are on the table, including all ideas we have not thought of yet. We merely say that with respect to these four mysteries of the cosmos  we lay prostrate, waiting for and working toward the day of revelation. In the meantime, in the cradle of these mysteries we humbly cannot elevate any as likely particle explanations within the domain that we have defined for the SM. 

In the case of dark matter, which is perhaps the most concrete cosmological mystery to solve, the resolution might be primordial black holes (PBHs)~\cite{Carr:2019yxo}, which do not require in and of themselves to extend the SM. However, new particles that give rise to special inflationary potentials that produce the right mass spectrum of PBHs might be required\footnote{See, for example, sec. IV of~\cite{Orlofsky:2016vbd} for a review of possibilities.}, and the two mysteries of inflation and dark matter may become intertwined~\cite{Motohashi:2017kbs,Ballesteros:2017fsr}. Perhaps even baryogenesis arises from reheating dynamics after inflation, and then three mysteries are intertwined. Or perhaps inflation occurs without the need for new particles, such as Higgs-inflation~\cite{Bezrukov:2007ep,Shaposhnikov:2015mja,Rubio:2018ogq,Drees:2019xpp}. We do not know yet. The point is that they are all mysteries at this stage in the sense that not one idea is compelling over other ideas.

What are the practical applications of attaching the ``four mysteries of the cosmos" to the SM definition? Is this attitude an abdication of scientific explanation?  No, it is not abdication of scientific pursuit. Its practical application is to put into the BSM column {\it any articulated} concrete idea that explains any one of the four mysteries. No concrete explanation for the mysteries can be part of the SM. The SM accepts the mystery. The SM is the mystery. That is  yet another reason why a scientist is not content with the SM, and BSM theories must be pursued. 

One might worry that accepting the four mysteries of the cosmos as an integral part of the SM definition means that the SM cannot be falsified from any cosmological data now or into the deep far future. That is correct, but it does not mean a new and better SM will not be found. No result of CMB measurements, dark matter distributions, etc., is anticipated that could ever show that the current SM under this definition is wrong. However, one hopes the day will come where enough data is accrued and enough theoretical insight is achieved to produce an articulated theory, which is not the SM (i.e., being that it would have less ``mysteries"), that explains the data efficiently and compellingly. That is the day a better and more refined SM would be born, even though the old SM would remain  compatible with all the data since by definition it took no concrete stand\footnote{Analogously, one  recalls that the SM of much of the western world in 325 A.D., as expressed by the Council of Nicaea, held that ``God [is] maker of all things both seen and unseen"~\cite{Tanner:1990}. SM-325 remains compatible with all the data, but as a theory it has been continually augmented over the years by articulated, computable, and proximate explanatory theories for the mysteries of natural phenomena.}.

If such a new compelling and concrete SM can be defined with respect to cosmological evolution, it will likely come complete with new particles and new interactions. Previously, Davoudiasl et al.~\cite{Davoudiasl:2004be} bravely made concrete choices (canonized ``myths" in our language) to explain at least three of the mysteries. There were many who agreed that these choices were among a small set of leading choices of the day, and perhaps it was a legitimate definition of a full SM in 2004 (SM-2004). However, the DM explanation is increasingly strained by LHC and by WIMP DM searches, and its status is very much reduced in many researcher's minds. Furthermore, the simple $m^2\phi^2$ inflationary potential is more or less ruled out now by CMB data\footnote{See, for example, fig.~8 of~\cite{Akrami:2018odb}.}. The model would not be put forward today as the SM choice. 

At a previous time the community's dominant preference for DM was weakly interacting massive particle (WIMP) near the weak scale. The early days of supersymmetry, especially, gave ascendancy to this DM candidate since it fell out of the supersymmetric spectrum ``for free"~\cite{Jungman:1995df}. However, intrusive searches for WIMP DM have been coming up empty for several decades now~\cite{Aprile:2017iyp,Akerib:2016vxi,Ackermann:2015zua}. It is certainly not ruled out, but the pressure on the idea has intensified. So what was once probably thought to be the standard theory explanation for DM now has strong additional competitors~\cite{Schirber:2018}. A major competitor is the axion, which not only can serve as DM~\cite{Abbott:1982af,Preskill:1982cy}, but  was invented originally to solve the different problem of suppressing perceived strong sector sources of CP violations~\cite{Peccei:1977hh,Hook:2018dlk}. Much work still must go into finding the axion or closing its full window of parameter space. One might be tempted to consider as part of SM-2019 a simple axion model of DM, like~\cite{Ballesteros:2016xej} has done. But the concern is that it too will be viewed in time as merely the next popular idea for all of DM in a long line of others that were not terribly compelling in absolute terms.

The SM today, it is my claim, should accept the four mysteries of the cosmos, and strive for the day, through experimental and theory work, when the SM no longer looks attractive in the face of a more concrete and compelling BSM theory, which then becomes the new SM with less mystery.

\xsection{Summary \& Conclusion}

In summary, we have argued that the simplest and most conservative (i.e., the least ``new" experimental consequences demanded from it) definition of the Standard Model that is adequate for today (SM-2019) is a theory that simultaneously holds the following facts (quarks, leptons and their interactions), mysteries (cosmological mysteries) and myths (neutrino sector):
\begin{itemize}
\item  Gauge symmetries are $SU(3)_c\times SU(2)_L\times U(1)_Y$, with no additional discrete mod-ing (such as $Z_2$, $Z_3$, or $Z_6$), with their accompanying gauge bosons.
\item  $SU(2)_L\times U(1)_Y\to U(1)_{\rm em}$ is accomplished by a single Higgs doublet, which gives mass to the $W^\pm$ and $Z^0$ bosons, and manifests a single propagating elementary scalar, the Higgs boson ($h^0$).
\item The elementary particle content has the gauge bosons (photon, gluons and $W^\pm/Z^0$ weak bosons),  three generations of fermions (quarks and leptons), and the Higgs boson. Both left- and right-handed neutrinos are present in the spectrum.
\item $B-L$ is a conserved global quantum number, which forbids $\nu_R$ Majorana masses.
\item $\theta_{\rm QCD}=0$.
\item In the limit of zero gravity, there are only renormalizable interactions among the elementary particles listed above, and all of those interactions must be consistent with the above symmetries and with Poincar\'e space-time symmetry.
\item The neutrino masses are entirely Dirac masses (implication of above conditions), and their masses obey the normal hierarchy (NH) solution.
\item The ``four mysteries of the cosmos" (cosmological constant, dark matter, inflation, baryogenesis) are accepted as mysteries of the SM without concrete demands for new elementary particles or interactions.  
\end{itemize}
Few would bet their lives on the validity of every line of the SM definition given above, nor would they {\it on any other} precise formulation that could have been offered. That is what makes particle physics perpetually susceptible to revolution. There are presently many BSM theories that challenge the primacy and stability of the SM formulation above, and there are many current experiments and proposed experiments looking for corresponding new states and new interactions.  Nevertheless, the above definition of the SM is presently self-consistent, satisfies all known data, and presents at least as economical structure to explain data and make predictions as any other postulated theory. It can be used to track our progress.

Let us not forget that particle physics has changed rather decisively since the early 1970s, but we have retained the name ``Standard Model" throughout it all. This confusion on what exactly is the SM has led some people less versed in the history of particle physics to think that nothing has changed because there is no new name. And, it has led to yet another group of people on the opposite end declaring that we are in a permanent state of beyond the SM because neutrinos have mass. Not recognizing substantial, albeit slow, progress in our evolution of what constitutes the standard theory of elementary particle physics has even had the implicit effect of hypnotizing some into thinking that no noteworthy progress will ever come until spacetime is totally upended and revolutionized through fermionic extra dimensions (supersymmetry), bosonic  extra dimensions (Randall-Sundrum, etc.), or manifestations of string excitations. This desensitization to more modest scientific progress, which is relentless yet often not totally surprising when the day of confirmation finally arrives, is connected with diffusing the SM name into a fog across a continent of technically different theories.

This article has discussed what it means to have a standard theory of elementary particle physics, and it has attempted to motivate the value of being precise about what our standard theory really is at every given moment and perhaps even having labels that change when the standard theory changes. In that sense, the SM of 1974 is different than the SM of 1988, which is different than the SM of other years, and so on, until we reach the SM of today:  SM-2019. I believe it is a mistake to think the SM can be just a private choice and is not worth articulating more precisely in a single coherent position. Each of us has heard  many private choices that most would not find compelling. Even among the experts there are many who would have liked some time ago for minimal $SU(5)$ GUTs and proton decay operators to be part of the standard theory~\cite{Bjorken:1984}. Others joked in the mid-1990s that the SM was really the minimal supersymmetric standard model (MSSM). I have tried to give here the most minimal and conservative definition to SM-2019, including identifying its mysteries and proposing a useful myth for the neutrino sector, based not on what I think but on what I think the community's sentiments could plausibly agree to. 

Our standard theory should be articulated often in order to set unambiguous targets for future work and, just as importantly, to track over time our changes of outlook, improvements in understanding, and gains in knowledge.  And that is how even ``modest" progress can be recognized for what it is: progress.

\medskip\noindent
{\it Acknowledgments:}   Support provided in part by the DOE (DE-SC0007859). I wish to thank G.~Ballesteros and A.~Pierce for helpful discussions.


\begin{thebibliography}{99}

\bibitem{Abazajian:2016yjj} 
  K.~N.~Abazajian {\it et al.} [CMB-S4 Collaboration],
  ``CMB-S4 Science Book, First Edition,''
  arXiv:1610.02743 [astro-ph.CO].
  
\bibitem{Abi:2018dnh} 
  B.~Abi {\it et al.} [DUNE Collaboration],
  ``The DUNE Far Detector Interim Design Report Volume 1: Physics, Technology and Strategies,''
  arXiv:1807.10334 [physics.ins-det].
 
\bibitem{Abbott:1982af} 
  L.~F.~Abbott and P.~Sikivie,
  ``A Cosmological Bound on the Invisible Axion,''
  Phys.\ Lett.\ B {\bf 120}, 133 (1983)
  [Phys.\ Lett.\  {\bf 120B}, 133 (1983)].
  doi:10.1016/0370-2693(83)90638-X
  
 
\bibitem{Ackermann:2015zua} 
  M.~Ackermann {\it et al.} [Fermi-LAT Collaboration],
  ``Searching for Dark Matter Annihilation from Milky Way Dwarf Spheroidal Galaxies with Six Years of Fermi Large Area Telescope Data,''
  Phys.\ Rev.\ Lett.\  {\bf 115}, no. 23, 231301 (2015)
  doi:10.1103/PhysRevLett.115.231301
  [arXiv:1503.02641 [astro-ph.HE]].
  
  
\bibitem{Agostini:2017jim} 
  M.~Agostini, G.~Benato and J.~Detwiler,
  ``Discovery probability of next-generation neutrinoless double-beta decay experiments,''
  Phys.\ Rev.\ D {\bf 96}, no. 5, 053001 (2017)
  doi:10.1103/PhysRevD.96.053001
  [arXiv:1705.02996 [hep-ex]].
  
\bibitem{Akerib:2016vxi} 
  D.~S.~Akerib {\it et al.} [LUX Collaboration],
  ``Results from a search for dark matter in the complete LUX exposure,''
  Phys.\ Rev.\ Lett.\  {\bf 118}, no. 2, 021303 (2017)
  doi:10.1103/PhysRevLett.118.021303
  [arXiv:1608.07648 [astro-ph.CO]].
  
\bibitem{Akrami:2018odb} 
  Y.~Akrami {\it et al.} [Planck Collaboration],
  ``Planck 2018 results. X. Constraints on inflation,''
  arXiv:1807.06211 [astro-ph.CO].
  
  
\bibitem{Altarelli:2010gt} 
  G.~Altarelli and F.~Feruglio,
  ``Discrete Flavor Symmetries and Models of Neutrino Mixing,''
  Rev.\ Mod.\ Phys.\  {\bf 82}, 2701 (2010)
  doi:10.1103/RevModPhys.82.2701
  [arXiv:1002.0211 [hep-ph]].
  
\bibitem{Aprile:2017iyp} 
  E.~Aprile {\it et al.} [XENON Collaboration],
  ``First Dark Matter Search Results from the XENON1T Experiment,''
  Phys.\ Rev.\ Lett.\  {\bf 119}, no. 18, 181301 (2017)
  doi:10.1103/PhysRevLett.119.181301
  [arXiv:1705.06655 [astro-ph.CO]].
  
\bibitem{Ballesteros:2016xej} 
  G.~Ballesteros, J.~Redondo, A.~Ringwald and C.~Tamarit,
  ``Standard Model-axion-seesaw-Higgs portal inflation. Five problems of particle physics and cosmology solved in one stroke,''
  JCAP {\bf 1708}, no. 08, 001 (2017)
  doi:10.1088/1475-7516/2017/08/001
  [arXiv:1610.01639 [hep-ph]].
  
\bibitem{Ballesteros:2017fsr} 
  G.~Ballesteros and M.~Taoso,
  ``Primordial black hole dark matter from single field inflation,''
  Phys.\ Rev.\ D {\bf 97}, no. 2, 023501 (2018)
  doi:10.1103/PhysRevD.97.023501
  [arXiv:1709.05565 [hep-ph]].
  
  
\bibitem{Banks:2010zn} 
  T.~Banks and N.~Seiberg,
  ``Symmetries and Strings in Field Theory and Gravity,''
  Phys.\ Rev.\ D {\bf 83}, 084019 (2011)
  doi:10.1103/PhysRevD.83.084019
  [arXiv:1011.5120 [hep-th]].
  
\bibitem{Bezrukov:2007ep} 
  F.~L.~Bezrukov and M.~Shaposhnikov,
  ``The Standard Model Higgs boson as the inflaton,''
  Phys.\ Lett.\ B {\bf 659}, 703 (2008)
  doi:10.1016/j.physletb.2007.11.072
  [arXiv:0710.3755 [hep-th]].
  
\bibitem{Bilenky:1977du} 
  S.~M.~Bilenky, S.~T.~Petcov and B.~Pontecorvo,
  ``Lepton Mixing, mu --> e + gamma Decay and Neutrino Oscillations,''
  Phys.\ Lett.\  {\bf 67B}, 309 (1977).
  doi:10.1016/0370-2693(77)90379-3
  
  \bibitem{Bjorken:1984}
  J.D.~Bjorken. ``The November Revolution: A theorist reminisces." Presented at the SLAC Symposium on the Tenth Anniversary of the November Revolution, Stanford, California, November 14, 1984 (Fermilab-Conf-85/58).
  
\bibitem{Caldwell:2017mqu} 
  A.~Caldwell, A.~Merle, O.~Schulz and M.~Totzauer,
  ``Global Bayesian analysis of neutrino mass data,''
  Phys.\ Rev.\ D {\bf 96}, no. 7, 073001 (2017)
  doi:10.1103/PhysRevD.96.073001
  [arXiv:1705.01945 [hep-ph]].
  
\bibitem{Carr:2019yxo} 
  B.~Carr,
  ``Primordial black holes as dark matter and generators of cosmic structure,''
  arXiv:1901.07803 [astro-ph.CO].
  
\bibitem{Chen:2004rr} 
  S.~L.~Chen, M.~Frigerio and E.~Ma,
  ``Large neutrino mixing and normal mass hierarchy: A Discrete understanding,''
  Phys.\ Rev.\ D {\bf 70}, 073008 (2004)
  Erratum: [Phys.\ Rev.\ D {\bf 70}, 079905 (2004)]
  doi:10.1103/PhysRevD.70.079905, 10.1103/PhysRevD.70.073008
  [hep-ph/0404084].
  
  
\bibitem{ChengLi:1984}
  T.~P.~Cheng and L.~F.~Li,
  {\it Gauge Theory Of Elementary Particle Physics.}\
  Oxford, UK: Clarendon, 1984.
 
\bibitem{Close:2004mj} 
  F.~Close,
  ``A November revolution: The birth of a new particle,''
  CERN Cour.\  {44N10}, 25 (2004).
  
\bibitem{Davis:1994jw} 
  R.~Davis,
 ``A review of the Homestake solar neutrino experiment,''
  Prog.\ Part.\ Nucl.\ Phys.\  {\bf 32}, 13 (1994).
  doi:10.1016/0146-6410(94)90004-3
  
  \bibitem{Davis:2002}
  R.~Davis. ``Raymond Davis Jr. Biographical." Autobiographical Sketch for Nobel Prize of 2002. 
  https://www.nobelprize.org/prizes/physics/2002/davis/biographical/ (accessed July 8, 2019).
  
\bibitem{Davoudiasl:2004be} 
  H.~Davoudiasl, R.~Kitano, T.~Li and H.~Murayama,
  ``The New minimal standard model,''
  Phys.\ Lett.\ B {\bf 609}, 117 (2005)
  doi:10.1016/j.physletb.2005.01.026
  [hep-ph/0405097].
  
\bibitem{DellOro:2016tmg} 
  S.~Dell'Oro, S.~Marcocci, M.~Viel and F.~Vissani,
  ``Neutrinoless double beta decay: 2015 review,''
  Adv.\ High Energy Phys.\  {\bf 2016}, 2162659 (2016)
  doi:10.1155/2016/2162659
  [arXiv:1601.07512 [hep-ph]].
  
\bibitem{Drees:2019xpp} 
  M.~Drees and Y.~Xu,
  ``Critical Higgs Inflation and Second Order Gravitational Wave Signatures,''
  arXiv:1905.13581 [hep-ph].
  
\bibitem{Ezquiaga:2017fvi} 
  J.~M.~Ezquiaga, J.~Garcia-Bellido and E.~Ruiz Morales,
  ``Primordial Black Hole production in Critical Higgs Inflation,''
  Phys.\ Lett.\ B {\bf 776}, 345 (2018)
  doi:10.1016/j.physletb.2017.11.039
  [arXiv:1705.04861 [astro-ph.CO]].
  
  
  
\bibitem{Fogli:2004ff} 
  G.~L.~Fogli, E.~Lisi, A.~Mirizzi and D.~Montanino,
  ``Probing supernova shock waves and neutrino flavor transitions in next-generation water-Cerenkov detectors,''
  JCAP {\bf 0504}, 002 (2005)
  doi:10.1088/1475-7516/2005/04/002
  [hep-ph/0412046].
  
\bibitem{Fukuda:1998mi} 
  Y.~Fukuda {\it et al.} [Super-Kamiokande Collaboration],
  ``Evidence for oscillation of atmospheric neutrinos,''
  Phys.\ Rev.\ Lett.\  {\bf 81}, 1562 (1998)
  doi:10.1103/PhysRevLett.81.1562
  [hep-ex/9807003].
 
\bibitem{Gariazzo:2018pei} 
  S.~Gariazzo, M.~Archidiacono, P.~F.~de Salas, O.~Mena, C.~A.~Ternes and M.~Tórtola,
  ``Neutrino masses and their ordering: Global Data, Priors and Models,''
  JCAP {\bf 1803}, no. 03, 011 (2018)
  doi:10.1088/1475-7516/2018/03/011
  [arXiv:1801.04946 [hep-ph]].
  
\bibitem{GellMann:1980vs} 
  M.~Gell-Mann, P.~Ramond and R.~Slansky,
  ``Complex Spinors and Unified Theories,''
  Conf.\ Proc.\ C {\bf 790927}, 315 (1979)
  [arXiv:1306.4669 [hep-th]].

\bibitem{Glashow:1961tr} 
  S.~L.~Glashow,
  ``Partial Symmetries of Weak Interactions,''
  Nucl.\ Phys.\  {\bf 22}, 579 (1961).
  doi:10.1016/0029-5582(61)90469-2
  
\bibitem{Harlow:2018tng} 
  D.~Harlow and H.~Ooguri,
  ``Symmetries in quantum field theory and quantum gravity,''
  arXiv:1810.05338 [hep-th].
  
    
\bibitem{Heavens:2018adv} 
  A.~F.~Heavens and E.~Sellentin,
  ``Objective Bayesian analysis of neutrino masses and hierarchy,''
  JCAP {\bf 1804}, no. 04, 047 (2018)
  doi:10.1088/1475-7516/2018/04/047
  [arXiv:1802.09450 [astro-ph.CO]].
  
  \bibitem{Hill:2019}
  H.~Hill, ``Lower limit on the half-life of neutrinoless double-beta decay." Phys.\ Today (15 Oct 2019) DOI:10.1063/PT.6.1.20191015a.
  
\bibitem{Hook:2018dlk} 
  A.~Hook,
  ``TASI Lectures on the Strong CP Problem and Axions,''
  arXiv:1812.02669 [hep-ph].
  
\bibitem{Jungman:1995df} 
  G.~Jungman, M.~Kamionkowski and K.~Griest,
  ``Supersymmetric dark matter,''
  Phys.\ Rept.\  {\bf 267}, 195 (1996)
  doi:10.1016/0370-1573(95)00058-5
  [hep-ph/9506380].
  
\bibitem{Kobayashi:1973fv} 
  M.~Kobayashi and T.~Maskawa,
  ``CP Violation in the Renormalizable Theory of Weak Interaction,''
  Prog.\ Theor.\ Phys.\  {\bf 49}, 652 (1973).
  doi:10.1143/PTP.49.652
  
  \bibitem{Laudan:1991}
  L.~Laudan, J.~Leplin. ``Empirical equivalence and underdetermination." {\it J.\ Phil.}\ 88, 449 (1991).
  
\bibitem{Lee:1977qz} 
  B.~W.~Lee, S.~Pakvasa, R.~E.~Shrock and H.~Sugawara,
  ``Muon and Electron Number Nonconservation in a V-A Gauge Model,''
  Phys.\ Rev.\ Lett.\  {\bf 38}, 937 (1977)
  Erratum: [Phys.\ Rev.\ Lett.\  {\bf 38}, 1230 (1977)].
  doi:10.1103/PhysRevLett.38.937, 10.1103/PhysRevLett.38.1230
  
\bibitem{Lei:2019nma} 
  M.~Lei, N.~Steinberg and J.~D.~Wells,
  ``Probing Non-Standard Neutrino Interactions with Supernova Neutrinos at Hyper-K,''
  arXiv:1907.01059 [hep-ph].
  
\bibitem{Long:2017dru} 
  A.~J.~Long, M.~Raveri, W.~Hu and S.~Dodelson,
  ``Neutrino Mass Priors for Cosmology from Random Matrices,''
  Phys.\ Rev.\ D {\bf 97}, no. 4, 043510 (2018)
  doi:10.1103/PhysRevD.97.043510
  [arXiv:1711.08434 [astro-ph.CO]].
  
\bibitem{Marciano:1977wx} 
  W.~J.~Marciano and A.~I.~Sanda,
  ``Exotic Decays of the Muon and Heavy Leptons in Gauge Theories,''
  Phys.\ Lett.\  {\bf 67B}, 303 (1977).
  doi:10.1016/0370-2693(77)90377-X
  
  
\bibitem{Minkowski:1977sc} 
  P.~Minkowski,
  ``$\mu \to e\gamma$ at a Rate of One Out of $10^{9}$ Muon Decays?,''
  Phys.\ Lett.\  {\bf 67B}, 421 (1977).
  doi:10.1016/0370-2693(77)90435-X
  
\bibitem{Mohapatra:1979ia} 
  R.~N.~Mohapatra and G.~Senjanovic,
  ``Neutrino Mass and Spontaneous Parity Nonconservation,''
  Phys.\ Rev.\ Lett.\  {\bf 44}, 912 (1980).
  doi:10.1103/PhysRevLett.44.912
  
\bibitem{Motohashi:2017kbs} 
  H.~Motohashi and W.~Hu,
  ``Primordial Black Holes and Slow-Roll Violation,''
  Phys.\ Rev.\ D {\bf 96}, no. 6, 063503 (2017)
  doi:10.1103/PhysRevD.96.063503
  [arXiv:1706.06784 [astro-ph.CO]].
    
\bibitem{Orlofsky:2016vbd} 
  N.~Orlofsky, A.~Pierce and J.~D.~Wells,
  ``Inflationary theory and pulsar timing investigations of primordial black holes and gravitational waves,''
  Phys.\ Rev.\ D {\bf 95}, no. 6, 063518 (2017)
  doi:10.1103/PhysRevD.95.063518
  [arXiv:1612.05279 [astro-ph.CO]].
  
  
  \bibitem{PDG:1992}
  Particle Data Group (K. Hikasa et al.).\ {``Review of Particle Properties.''} Phys.\ Rev.\ D45, S1 (1992). 
  
  \bibitem{PDG:1994}
  Particle Data Group (L.~Montanet et al.).\ {``Review of Particle Properties.''} Phys.\ Rev.\ D50, 1173 (1994). 
  
\bibitem{Peccei:1977hh} 
  R.~D.~Peccei and H.~R.~Quinn,
  ``CP Conservation in the Presence of Instantons,''
  Phys.\ Rev.\ Lett.\  {\bf 38}, 1440 (1977).
  doi:10.1103/PhysRevLett.38.1440
  
  \bibitem{Pinch:1986}
  T.~Pinch. {\it Confronting Nature: The Sociology of Solar-Neutrino Detection.} Boston: D. Reidel, 1986.
  
\bibitem{Preskill:1982cy} 
  J.~Preskill, M.~B.~Wise and F.~Wilczek,
  ``Cosmology of the Invisible Axion,''
  Phys.\ Lett.\ B {\bf 120}, 127 (1983)
  [Phys.\ Lett.\  {\bf 120B}, 127 (1983)].
  doi:10.1016/0370-2693(83)90637-8
  
\bibitem{Rubio:2018ogq} 
  J.~Rubio,
  ``Higgs inflation,''
  Front.\ Astron.\ Space Sci.\  {\bf 5}, 50 (2019)
  doi:10.3389/fspas.2018.00050
  [arXiv:1807.02376 [hep-ph]].
  
  
\bibitem{Salam:1968rm} 
  A.~Salam,
  ``Weak and Electromagnetic Interactions,''
  Conf.\ Proc.\ C {680519}, 367 (1968).
  
\bibitem{Schechter:1980gr} 
  J.~Schechter and J.~W.~F.~Valle,
  ``Neutrino Masses in $SU(2)\times U(1)$ Theories,''
  Phys.\ Rev.\ D {\bf 22}, 2227 (1980).
  doi:10.1103/PhysRevD.22.2227
  
  \bibitem{Schirber:2018}
  M.~Schirber. ``WIMP Alternatives Come Out of the Shadows."  APS Physics. https://physics.aps.org/articles/v11/48 (accessed July 8, 2019).
  
  
\bibitem{Schwetz:2017fey} 
  T.~Schwetz, K.~Freese, M.~Gerbino, E.~Giusarma, S.~Hannestad, M.~Lattanzi, O.~Mena and S.~Vagnozzi,
  ``Comment on "Strong Evidence for the Normal Neutrino Hierarchy",''
  arXiv:1703.04585 [astro-ph.CO].
  
\bibitem{Shaposhnikov:2015mja} 
  M.~Shaposhnikov,
  ``The Higgs boson and cosmology,''
  Phil.\ Trans.\ Roy.\ Soc.\ Lond.\ A {\bf 373}, no. 2032, 20140038 (2015).
  doi:10.1098/rsta.2014.0038
  
\bibitem{Simpson:2017qvj} 
  F.~Simpson, R.~Jimenez, C.~Pena-Garay and L.~Verde,
  ``Strong Bayesian Evidence for the Normal Neutrino Hierarchy,''
  JCAP {\bf 1706}, no. 06, 029 (2017)
  doi:10.1088/1475-7516/2017/06/029
  [arXiv:1703.03425 [astro-ph.CO]].
  
  \bibitem{Tanner:1990}
  N.P.~Tanner, {\it Decrees of the Ecumenical Councils} (2 vols.)\ Georgetown University Press, 1990.
  
\bibitem{Trimble:1973ca} 
  V.~Trimble and F.~Reines,
  ``The solar neutrino problem - a progress(?) report,''
  Rev.\ Mod.\ Phys.\  {\bf 45}, 1 (1973).
  doi:10.1103/RevModPhys.45.1
  
  
\bibitem{Valle:2017fwa} 
  J.~W.~F.~Valle,
  ``Neutrino physics from A to Z : two lectures at Corfu,''
  PoS CORFU {\bf 2016}, 007 (2017)
  doi:10.22323/1.292.0007
  [arXiv:1705.00872 [hep-ph]].
  
\bibitem{Weinberg:1967tq} 
  S.~Weinberg,
  ``A Model of Leptons,''
  Phys.\ Rev.\ Lett.\  {\bf 19}, 1264 (1967).
  doi:10.1103/PhysRevLett.19.1264
  
\bibitem{Wells:2018nwj} 
  J.~D.~Wells,
  ``Beyond the hypothesis: Theory's role in the genesis, opposition, and pursuit of the Higgs boson,''
  Stud.\ Hist.\ Phil.\ Sci.\ B {\bf 62}, 36 (2018).
  doi:10.1016/j.shpsb.2017.05.004
  
\bibitem{Wells:2019zrj} 
  J.~D.~Wells,
  ``Discovery goals and opportunities in high energy physics: a defense of BSM-oriented exploration over signalism,''
  arXiv:1904.02769 [physics.hist-ph].
  
\bibitem{Winter:2010hb} 
  W.~Winter,
  ``Lectures on neutrino phenomenology,''
  Nucl.\ Phys.\ Proc.\ Suppl.\  {203-204}, 45 (2010)
  doi:10.1016/j.nuclphysbps.2010.08.005
  [arXiv:1004.4160 [hep-ph]].
  
\bibitem{Yanagida:1979as} 
  T.~Yanagida,
  ``Horizontal gauge symmetry and masses of neutrinos,''
  Conf.\ Proc.\ C {7902131}, 95 (1979).
  
  
\end{thebibliography}
\end{document}